\documentclass[12pt]{article}
\usepackage{epsfig}

\parskip=\medskipamount 
plus.3em minus.5em \abovedisplayshortskip=.5em plus.2em minus.4em
\renewcommand{\thesection}{\arabic{section}}

\catcode`@=11

\@addtoreset{equation}{section} \@addtoreset{equation}{subsection}
\def\theequation{\ifnum\value{section}=0 \arabic{equation}\ignorespaces
\else \ifnum\value{section}=-1 A.\arabic{equation}\ignorespaces
\else \ifnum\value{subsection}=0
\thesection.\arabic{equation}\ignorespaces \else
\thesection.\arabic{subsection}.\arabic{equation}\ignorespaces
                             \fi
                        \fi
                   \fi}

{\catcode`\'=\active \def'{{}^\bgroup\prim@s}}

\catcode`@=12

\newcommand{\bq}{\begin{equation}}
\newcommand{\be}{\begin{equation}}
\newcommand{\fq}{\end{equation}}
\newcommand{\ee}{\end{equation}}
\newcommand{\bqr}{\begin{eqnarray}}
\newcommand{\beqs}{\begin{eqnarray}}
\newcommand{\fqr}{\end{eqnarray}}
\newcommand{\eeqs}{\end{eqnarray}}

\newcommand{\rf}[1]{(\ref{#1})}

\def\bop#1{\setbox0=\hbox{$#1M$}\mkern1.5mu
    \vbox{\hrule height0pt depth.04\ht0
    \hbox{\vrule width.04\ht0 height.9\ht0 \kern.9\ht0
    \vrule width.04\ht0}\hrule height.04\ht0}\mkern1.5mu}

\begin{document}
\thispagestyle{empty}

\begin{flushright}
\begin{tabular}{l}
hep-th/0503062 \\
\end{tabular}
\end{flushright}

\vskip .6in
\begin{center}

{\bf  Derivation of Quantum Field Dynamics}

\vskip .6in

{\bf Gordon Chalmers}
\\[5mm]

{e-mail: gordon@quartz.shango.com}

\vskip .5in minus .2in

{\bf Abstract}

\end{center}

The perturbative dynamics of quantum field theories is described by
a recursive expansion similar to the well known loop expansion.
The equivalent formulation based on low-energy dynamics via an
expansion in derivatives is well known in the literature; this
is described by terms from low energy to high energy.  The
coefficients of these terms are presented in a simplified algebraic
manner.  In general, the dynamics of any bare Lagrangian, including
additional higher dimension terms, is found by iteration in a
discrete algorithm.  Inversion of the quantized dynamics to the
fundamental theory is possible.

\vfill\break

The dynamics of quantum field theories in general
is typically examined within the loop expansion.  The integrals are
complicated, and the symmetries of the theory are not always
manifest in this formalism.  For example, the well known low
energy solution in two derivatives of supersymmetric field and
string theories can be examined via holomorphy, which is useful
for the understanding of phase transitions.  The approach in
derivatives extends this approach to higher energies, various
regimes in coupling, and shows relations between seemingly diverse
theories \cite{ChalmersUnPub1}-\cite{Chalmers8}.

The iterative formalism in \cite{ChalmersUnPub1}-\cite{Chalmers8}
is examined here in more detail and simplified with respect to the
derivation of the coefficients.  The coefficients are generally
derived for models in particle physics, condensed matter models,
and partially in string theories (in which a question regarding the
coefficients that describe the corners of moduli space remain).

A solution to the coefficients of the derivative and coupling
expansion allows a determination of relations between them and to
coefficients in other theories including holography, weak-weak
dualities, and strong-weak dualities.  The accurate modeling of
experimental phenomena requires the accurate determination of these
coefficients; in principle, the quantization of a bare action with
the inclusion of higher dimension operators (and string fields)
allows for a perfect match with experimental phenomena.

Furthermore, the determination of the coefficients of the
effective action, without integrals, but solely in terms of
hidden symmetries will permit an extension of the determination
to more complicated theories, such as strings in curved backgrounds;
the calculations in these examples requires terms in the effective
action containing spin degrees of freedom of arbitrarily high
number.

Consider for example, the prototype scalar field theory described
by

\bqr
{\cal L} = {1\over 2} \phi \partial^\mu\partial_\mu \phi + {1\over
2} m^2 \phi^2 + {\lambda_3\over 3!} \phi^3 + {\lambda_4\over 4!}
\phi^4 \ .
\label{scalartheory}
\fqr
The inclusion of possible higher derivative terms,

\bqr
{\cal L}' = {\lambda_6\over 6!} {\phi^6\over \Lambda^2} + \ldots
\label{improvedtheory}
\fqr
may be placed in the initial Lagrangian.  This inclusion is
typically examined in the renormalization group flow, but also has
a consequence in the modeling of the phenomena as found for
example in experiment (e.g. condensed matter or particle dynamics
in high energy theory including the mass derivation).  Mathematically
these terms are of also interest.

The bare theory in \rf{scalartheory} and \rf{improvedtheory} may
be examined in the usual quantum expansion.  The 'tree-level' or
classical vertices are the usual Feynman rules.  However, the
derivative expansion vertices are derived via expanding the
classical scattering.  These terms are defined by the kinematic
invariants as,

\bqr
\lambda_{n}^{(p_{11},p_{12},\ldots,p_{nn})} =
\lambda_{n,0}^{(p_{11},p_{12},\ldots,p_{nn})} \prod s_{ij}^{p_{ij}}
\label{momentumvertices}
\fqr
through the series of numbers $n$ and $p_{ij}$.  The case of $p_{ij}=
p_{00}$ is a scale $\Lambda^{p_{00}}$.  In the case of multiple
masses, a cutoff, or a gravitational scale, the zero index
$p_{00}$ is further indexed to $p_{00,a}$.  The redundancy
in $p_{ij}$ via momentum conservation is not included.  The
non-analytic terms required by unitarity, that is, the logarithms
such as $\ln s_{ij}$ are constructed via the usual perturbative
unitarity relation ${\rm Im} S= S^\dagger S$.  These terms are
not examined in this work.

These on-shell vertices are, for the mentioned scalar field theory,
\bqr
\lambda_3^{(0)} = \lambda_3 \quad\quad
\lambda_4^{(0)} = \lambda_3^2/m^2 + \lambda_4
\cr
\lambda_4^{(n,0,\ldots,0)} = \lambda_3^2 (-s_{12})^n/m^{2+2n} \ ,
\fqr
for example.  The general vertex is found via the expansion of the
classical Feynman graphs, or the classical scattering.

The full vertex of the prescribed momenta terms including the
coupling expansions of a $\phi^4$ theory,

\bqr
g_n^{p_{11},p_{12},\ldots, p_{nn}} = \sum_{g=0}^\infty
\lambda^{n-1+g} \alpha^{p_{11},p_{12},\ldots,p_{nn}}_{n,g} \ ,
\label{expansion}
\fqr
with the $a$ parameters rational numbers.  Due to momentum
conservation there are relations between the parameters
$a_{p_{11}, p_{12}, \ldots, p_{nn}}$.

The fundamental iteration is accomplished via the sewing procedure
as described in \cite{Chalmers1}-\cite{Chalmers8}.  The integrals
are simple free-field ones in x-space, and generate an infinite
series of relations between the parameters of the coupling
expansion $\alpha^{p_{ij}}_{n,g}$ in \rf{expansion}.  These
relations in theories with the symmetry $\phi$ to $-\phi$
are,

\bqr
\sum_q^{p_{ij}} \alpha^{p_{ij}}_{n,q} \lambda^{(n-2)/2 + q} =
 \sum_{i,j,p;l_{ij},n_{ij},m_{ij}} \alpha_{n+p,i}^{l_{ij}}
\alpha_{n+p,j}^{n_{ij}} \lambda^{n+p+i+j-2}
  I_{l_{ij},n_{ij}}^{p_{ij}} \ .
\label{phiiteration}
\fqr
The indices $i,j$ are exampled below. The coefficients
$I_{l_{ij},m_{ij}}^{p_{ij}}$ are defined by the momentum expansion
of the integrals

\bqr
J_{{\tilde l}_{ij},{\tilde m}_{ij}}^{\sigma,{\tilde p}_{ij}} =
 \int d^dk \prod_{a=1}^p {1\over
(k - k_{\sigma(a)})^2 + m^2} \prod
 s_{ij}^{{\tilde l}_{ij}+{\tilde m}_{ij}} \quad\vert^{p_{ij}}
\ ,
\label{integrals}
\fqr
with ${\tilde l}_{ij}$ and ${\tilde m}_{ij}$ parameterizing a subset
of the lines of the vertex lines which are contracted inside the
loop. The integrals are symmetrized over the the external lines
in the formula \rf{phiiteration}; there are $n_1$ and $n_2$ external
lines on each side of the graph and $b$ parameterizes a subset of these
numbers (e.g. $n_1=1,2,3,4$, $n_2=5,6,7,8$ and $b=3,4,5,6$; the
$l_{ij}$ and $m_{ij}$ parameterize the kinematics associated with the
exernal and internal lines.  The expansion of the integral in
\rf{integrals} in the momenta generate the coefficients $p_{ij}$;
the set of numbers $\sigma(a)$ parameterizing the subset of
numbers of the two vertices (forming an integral with $n$ external
lines; the numbers $\sigma(a)$ label numbers beyond the external
lines $1,\ldots, n_1$ and $n_1+1,\ldots,n$) is actually
irrelevant in the final result to the formula in \rf{phiiteration};
this property lends to a group theory interpretation of the final
result in terms of the coefficients
\bqr
I_{l_{ij},n_{ij}}^{\sigma,p_{ij}} \ ,
\fqr
after summing the permutations.  The numbers $i,j$ in $l_{ij}$ and
$n_{ij}$ span $1$ to $m$ (including internal lines) and those in
$p_{ij}$ span $1$ to $n$:

\bqr
l_{ij}=(l_{ij},0,\ldots,0,l_{ij})
\qquad n_{ij}=(0,\ldots 0,n_{ij},\ldots,n_{ij},
 n_{ij},\ldots,n_{ij}) \ ,
\fqr
and
\bqr
p_{ij}=(p_{ij},\ldots,p_{ij}) \ .
\fqr
This notation of $l_{ij}$, $m_{ij}$, and $p_{ij}$ is used to setup
a (pseudo-conformal) group theory interpretation of the scattering.

The details of the expansion of the integrals in \rf{integrals}
depend on the selection of the internal lines found via the
momenta of the vertices

\bqr
\lambda_{n}^{(p_{11},p_{12},\ldots,p_{nn})}
\label{momemtavertex}
\fqr
on either side of the double vertex graph.  Although the ${\tilde
l}_{ij}$, ${\tilde m}_{ij}$, and ${\tilde p}_{ij}$ depend on the
details of the contractions and sums of the lines of the
individual vertices, the actual coefficients of the iteration,
i.e. $I_{l_{ij},n_{ij}}^{p_{ij}}$, are functions only of the
vertex parameters.  The details of the expansion and the contractions
of the tensors in the integrals \rf{integrals} are parameterized by
$p_{ij}$, which label the momentum expansion of the integrals.
The coefficients $p_{ij}$ range from $0$ to $\infty$, in
accordance with the momentum expansion of the massive theory.

Although the coefficients $I_{l_{ij},n_{ij}}^{p_{ij}}$ arise from
the integral expansion, they also have a group theory description.
The dynamics of the expansion are dictated via these coefficients
for an arbitrary initial condition of the bare Lagrangian.

The iteration of the coefficients results in the simple
expression,

\bqr
 \alpha^{m_{ij}}_{n,q}  =
 \sum_{i,j,p; l_{ij}, n_{ij}} \alpha_{n+p,i}^{l_{ij}}
\alpha_{n+p,j}^{n_{ij}} I_{l_{ij},n_{ij}}^{m_{ij}} \ .
\label{coeffiteration}
\fqr
The sums are over the number of internal lines $p$, the powers
of the shared couplings $i$ and $j$ (in $\phi^4$),

\bqr
n+q=2(n+p)-1+i+j \qquad q=n+2p-1+i+j
\fqr
and the numbers of momenta
$l_{ij}$ and $n_{ij}$ (some of which are within the integral).
The parameters $m_{ij}$ label the external momenta, interpreted
group theoretically through the coefficient $I$.

As an example of the procedure, consider the lowest order term
$\alpha_{4;1}^{m_{ij}}$.  It is found via

\bqr
\alpha_{4;1}^{m_{ij}} = \sum_{l_{ij},n_{ij}}
  \alpha_{4;0}^{l_{ij}} \alpha_{4;0}^{n_{ij}}
    I_{l_{ij},n_{ij}}^{m_{ij}} \ ,
\fqr
with a summation of $l_{ij}$ and $n_{ij}$, for example through the
internal lines.  As an example,

\bqr
m=(c_1,0,0,0,0,c_2) \qquad l=(a_1,0,0,0,0,a_2)
\fqr
\bqr
n=(b_1,0,0,0,0,b_2)
\label{example}
\fqr
with $c_1$ and $c_2$ satisfying $a_i+b_i=c_i$ by dimensional analysis
in this four-point example with two internal lines.  The
four-point function entering into the right-hand side of the
equation is the tree-level vertex, and by dimensional analysis
the momenta of the vertices have to be conserved.  There are more
general momenta configurations than that in \rf{example} that should
be included in the summation.

In general the coefficient derivation of the theory is determined
by the iteration algebraically of the formula in \rf{iteration}.
The non-analytic terms, for unitarity reasons, are derivable from
these coefficients via ${\rm Im}S=S^\dagger S$.

The general spin and gauge content may be included by adding more
indices on the coefficients and in the iteration.  These indices
in gauge theory are the spin content and the momentum content; the
general term contains contractions of the spin fields (e.g. the
polarizations) with other spin fields and with momenta.  In
principle, the masslessness of the particles (e.g. in gauge
theory) is included by letting the indices $l_{ij}$, $m_{ij}$ and
$p_{ij}$ be negative.  The tensors and the integrals are more
complicated in this case.

The general contraction of the spin fields with momenta is
accomplished via spin $1/2$ fields, e.g.,

\bqr
\varepsilon_i \cdot k_j \psi^\alpha k_{\alpha}^{\dot\alpha}
\psi_{\dot\alpha} \ldots \ .
\fqr

The general gauge theory numerator contains contractions of
spin $0$, $1/2$, and $1$ terms, and is represented via traces
of terms containing the fermions and gauge bosons,

\bqr
{\rm tr} \quad (s_1 k)
\qquad
{\rm tr} \quad (s_{1/2} k s_1 k s_{1/2}) \ .
\fqr
The general term may be labeled group theoretically (the Lorentz
representations contracted with momenta), via the permutations

\bqr
(s_1, s_2, s_3, \ldots, s_n) \qquad ({\bar s}_1, {\bar s}_2,
  {\bar s}_3, \ldots, {\bar s}_n) \ ,
\fqr
together with the momenta,

\bqr
k_\sigma=(k_{\sigma(1)}, k_{\sigma(2)}, \ldots, k_{\sigma(n)}) \ ,
\cr
k_{\bar\sigma}=(k_{{\bar\sigma}(1)}, k_{{\bar\sigma}(2)},
 \ldots, k_{{\bar\sigma}(n)}) \ ,
\fqr
which are labeled via the series in $\sigma(i)$ and ${\bar\sigma}(i)$.
In the case of momenta in terms of the spin variables $s_i$, the
variable $s_i=0$ is used.  The contractions of the polarization
vectors $\varepsilon^\mu$ are split into the representation
$(1/2,1/2)$ via the indices $\varepsilon^\alpha$ and
$\varepsilon^{\dot\alpha}$ (this is useful in the spinor helicity
formalism).  The trace terms are formulated via two additional
vectors $t=(a_1,a_2,a_3,\ldots)$ and ${\bar t}=({\bar a}_1,{\bar
a}_2,{\bar a}_3,\ldots)$ that contract

These fields and momenta spanned by $s$, ${\bar s}$ and $k_\sigma$,
$k_{\bar\sigma}$ are contracted via the tensors $t$ and $\bar t$.
The fields $s_i$ are labeled by the momenta $k_i$, and are contracted
in a cyclic manner associated via the tensors $t$ and $\bar t$
(e.g. $t=(1,1,1,1,2,2,2,2)$ for two spinors at positions $1$ and
$4$ and $5$ and $8$).

In practice the spinor helicity basis is used to simplify
calculations, and a judicious basis of reference momenta is chosen
to simplify the end result.  In this case, instead of the fields,
the reference momenta and line factors are inserted in place of
the fields \cite{ManganoParke}.

The group theory representations on the propagating fields are
also contracted in general with multiple traces,

\bqr
(g_1,g_2,\ldots,g_n) \ldots (h_1,h_2,\ldots, h_n) \ ,
\label{grouptheory}
\fqr
with $g_i$ the representation and $h$ the trace term in a subleading
context; for example a term such as ${\rm Tr} (T_{a_1} T_{a_2} \ldots
T_{a_m}) {\rm Tr} (T_{a_{m+1}} \ldots T_{a_n})$ corresponds to
$h=(1,1,\ldots, 1, 2,\ldots 2)$.

In this manner the iteration takes the form,

\bqr
 \alpha^{m_{ij}}_{n,q;s_a,{\bar s}_{\bar a},g_b}  =
 \sum_{i,j,p; l_{ij}, n_{ij}}
   \alpha_{n+p,i;s^{(1)}_a,{\bar s}^{(1)}_{\bar a},g^{(1)}_b,l_{ij}}
\alpha_{n+p,j;s^{(2)}_a,{\bar s}^{(2)}_{\bar a},g^{(2)}_b,n_{ij}}
\cr
  I_{l_{ij},n_{ij};s,s^{(1)},s^{(2)};g,g^{(1)},g^{(2)};m_{ij}} \ .
\label{iteration}
\fqr
A direct calculation of the integrals generates this formula,
after taking the product of the two general vertices and summing
the internal lines.  However, a group theory representation of the
coefficients $I$ is more elegant.  The interpretation of the quantum
wavefunction overlaps, i.e. $I$, would permit a simpler
derivation, and also generalize to generic theories and possibly
string theories.  The coefficients $I$ are determined by 1) the
volume region in which the integrals are defined, and 2) via the
lattice structure pertinent to condensed matter models.

For practical purposes, the kinematics of the scattering requires
the specification of the incoming and outgoing states, i.e. the
helicity and four-vectors of the particle states such as
$\psi^\alpha(k) = k^\alpha$ and $\varepsilon(k)_\mu^\pm$. These
line factors are typically utilized with the gauge invariance of
the amplitudes via the spinor helicity method \cite{ManganoParke}.
The choice of the reference momenta associated with the
polarizations of the gauge bosons is important to simplify both the
calculations and the end result of the amplitudes.

It is possible that the best choice for an immediate
simplification of the terms in the on-shell effective action is
dictated by a function of the quantum numbers specifying the
terms in the derivative expansion.  Of course, the terms in the
previous pages should be grouped into gauge invariant combination.
Each gauge invariant set of terms that contributes to the
amplitude can be chosen with a separate set of reference momenta.
The simplification of the addition of the individual terms
requires momentum conservation and spinor inner product identities
(e.g. Fierz identities), which can be very complicated in general.

The simplification and minimization of the use of the possibly
large number of Fierz and momenta identities, both within the gauge
invariant combinations and between these sets, could be accomplished
by the appropriate choice of reference momenta.  There is clearly
an ideal choice of these auxiliary parameters, for immediate
compression of the end result of the amplitude at each order in the
coupling; this ideal choice is defined by the least number of these
identities required to obtain the most compact result.

The choice of the non-trivial reference momenta in these terms is
possibly associated with polynomial equations parameterized by the
quantum numbers of the individual terms, with solutions generating
the reference momenta choices; for example, a possibility is that
one polynomial could specify the polarizations $i$ and auxiliary
momenta $j$ via a series of rational number solutions
$\sigma(i)/\rho(i)$ via $P_{q_i}(x)=\prod (x-x_j)$ and
$x_j=\sigma(i)/\rho(j)$ (with $q_i$ the quantum numbers of the
terms in the effective action).  Given a set of choices of
reference momenta $(\sigma(j),\rho(j))$ for the terms
contributing to the amplitude at a given coupling order there is
a minimal number $N_{\sigma,\rho}$ of momenta conservation and
Fierz identities required to reach the most compact form (fewest
number of additions).

Because the recent work on the projective variety twistor generation of
the tree and one-loop amplitudes appears to result in more compact forms
of the (partial) amplitudes, there is a natural reason to suggest that
this polynomial equation(s) is related to the twistor geometry and
its instantons.

The interpretation of the polynomials $P_{q_i}=\prod (x-x_j)$ and
the numbers $N(\sigma(j),\rho(j))$ could have interesting geometric
and number theoretic interpretations attached to gauge theory
dynamics, including gravity.

Furthermore, the inversion of the prescribed quantum field
coefficients, as for example found in experiment, appears direct
via the iteration procedure.  There is by definition a pseudo-linear
relation (derived via the iteration) between the bare coefficients
$\lambda_j$ of the classical theory, including the higher derivative
interactions, and those of the quantum operators $g_i(\lambda_j)$.
The inversion of the formula $g_i(\lambda_j)$ to $\lambda_i$
requires solving an infinite number of coupled polynomial
equations; the solution of which is useful also in pure mathematics.
The group theory interpretation of the iteration is of use in this,
and potentially leads to a simple formulation of a linear relation
between quantum coefficients to bare coefficients including higher
derivative interactions.

The generalization to string theory is direct.  In this case,
there are an infinite number of string fields.  The integrals and
the analysis are the same, utilizing the tree-level scattering and
the same integrals.  The corners of the moduli space not included
in the 'field'-theory integrals are compensated for via a series
of higher dimension operators with 'corner-moduli space'
parameters; this appears complicated, but the coefficients are
potentially found via a 'symmetry'.  The tree-level scattering is
determined via the quantum numbers labeling the $\alpha$ parameters,
labeling the Lorentz and group theory (the spin, kinematics, and gauge
theory).

The computation of the integral coefficients $I_{q}$ is rather
direct and should shed light on hidden symmetries (infinite
dimensional of a twisted Virasoro type) of the form,

\bqr
I_{q_1} I_{q_2} = I_{q_3}
\fqr
of the theories.  The multiplicative product would also simplify
the calculations.

The quantization of branes and extended dimensional membranes
(with possible deformations to avoid a continuous spectrum) is
relevant to string theory and mathematics.  The coefficients of
the improved terms required to complete the corners of the moduli
space integration in the string target space action derivation show
the 2-dimensionality of the string propagation.  The same can be
said about membrane scattering, i.e. $d>1$ membranes,  and non-standard
string propagation such as non-critical, and the corners
of the $d>2$ world-volume moduli space.  The geometry and world-sheet
action is described by these coefficients, to which there might be a
guiding principle to the numbers for similar world-sheet actions.
Also, the general $d$-dimensional form might provide insight into
a potential classification of $d$-dimensional compact Riemann
surfaces.

\vfill\break

\end{document}